\documentclass[runningheads]{llncs}
\usepackage{graphicx}
\usepackage{amsfonts}
\usepackage{amsmath}
\usepackage{hyperref}
\usepackage{blindtext}
\begin{document}
\title{Learning-based Optimization of the Under-sampling Pattern in MRI }

%
%
\author{Cagla Deniz Bahadir\inst{1}\and
Adrian V. Dalca \inst{2,3} \and Mert R. Sabuncu\inst{1,4} }
%
%
\institute{Meinig School of Biomedical Engineering, Cornell University \and Martinos Center for Biomedical Imaging, Massachusetts General Hospital, HMS \and Computer Science and Artificial Intelligence Lab, MIT \and School of Electrical and Computer Engineering, Cornell University }

%
%
\maketitle              

\begin{abstract}
Acquisition of Magnetic Resonance Imaging (MRI) scans can be accelerated by under-sampling in k-space (i.e., the Fourier domain).

In this paper, we consider the problem of optimizing the sub-sampling pattern in a data-driven fashion.
Since the reconstruction model's performance depends on the sub-sampling pattern, we combine the two problems. 
For a given sparsity constraint, our method optimizes the sub-sampling pattern \textit{and} reconstruction model, using an end-to-end learning strategy.
Our algorithm learns from full-resolution data that are under-sampled retrospectively, yielding a sub-sampling pattern and reconstruction model that are customized to the type of images represented in the training data.
The proposed method, which we call LOUPE (Learning-based Optimization of the Under-sampling PattErn), was implemented by modifying a U-Net, a widely-used convolutional neural network architecture, that we append with the forward model that encodes the under-sampling process.
Our experiments with T1-weighted structural brain MRI scans show that the optimized sub-sampling pattern can yield significantly more accurate reconstructions compared to standard random uniform, variable density or equispaced under-sampling schemes. The code is made available at:~\url{https://github.com/cagladbahadir/LOUPE} .

\end{abstract}
\section{Introduction}
MRI is a non-invasive, versatile, and reliable imaging technique that has been around for decades. A central difficulty in MRI is the long scan times that reduce accessibility and increase costs.
A method to speed up MRI is parallel imaging that relies on simultaneous multi-coil data acquisition and thus has hardware requirements. Another widely used acceleration technique is Compressed Sensing (CS)~\cite{lustig2008compressed}, which does not demand changes in the MR hardware.

MRI measurements are spatial frequency transform coefficients, also known as k-space, and images are computed by solving the inverse Fourier transform that converts k-space data into the spatial domain. Medical images often exhibit considerable spatial regularity. For example, intensity values usually vary smoothly over space, except at a small number of boundary voxels. This regularity leads to redundancy in k-space and creates an opportunity for sampling below the Shannon-Nyquist rate~\cite{lustig2008compressed}. Several Cartesian and non-Cartesian under-sampling patterns have been proposed in the literature and are widely used in practice, such as Random Uniform~\cite{gamper2008compressed}, Variable Density~\cite{wang2010variable} and equispaced Cartesian~\cite{haldar2011compressed} with skipped lines.

A linear reconstruction of under-sampled k-space data (i.e., a direct inverse Fourier) yields aliasing artifacts, which are challenging to distinguish from real image features for regular sub-sampling patterns. Stochastic sub-sampling patterns, on the other hand, create noise-like artifacts that are relatively easier to remove ~\cite{lustig2008compressed}. The classical reconstruction strategy in CS involves regularized regression, where a non-convex objective function that includes a data fidelity term and a regularization term is optimized for a given set of measurements. The regularization term reflects our \textit{a priori} knowledge of regularity in natural images.
Common examples include sparsity-encouraging penalties such as L1-norm on wavelet coefficients and total variation~\cite{ma2008efficient}.

In regularized regression, optimization is achieved via iterative numerical strategies, such as gradient-based methods, which can be computationally demanding. Furthermore, the choice of the regularizer is often arbitrary and not optimized in a data-driven fashion. These drawbacks can be addressed using machine learning approaches, which enable the use of models that learn from data and facilitate very efficient and fast reconstructions.

\subsection{Machine Learning for Under-sampled Image Reconstruction}

Dictionary learning techniques~\cite{huang2014bayesian,qu2014magnetic,ravishankar2011mr} have been used to implement customized penalty terms in regularized regression-based reconstruction. A common strategy is to project the images (or patches) onto a ``sparsifying'' dictionary. Thus, a sparsity-inducing norm, such as L1, on the associated coefficients can be used as a regularizer. The drawback of such methods is that they still rely on iterative numerical optimization, which can be computationally expensive.

Recently, deep learning has been used to speed up and improve the quality of under-sampled MRI reconstructions~\cite{lee2017deep,mardani2017deep,quan2018compressed,sun2016deep,yang2018dagan}. These models are trained on data to learn to map under-sampled k-space measurements to image domain reconstructions. For a new data point, this computation is often non-iterative and achieved via a single forward pass through the ``anti-aliasing'' neural network, which is computationally efficient. However, these machine learning-based methods are typically optimized for a specific under-sampling pattern provided by the user. Furthermore, there are also techniques that are optimizing the sub-sampling patterns for given reconstruction methods ~\cite{gozcu2018learning,baldassarre2016learning,mahabadi2019learning,mahabadi2018real}. The reconstruction model's performance will depend significantly on the sub-sampling pattern. In this paper, we are interested in optimizing the sub-sampling pattern in a data-driven fashion. Therefore, our method optimizes the sub-sampling pattern \textit{and} reconstruction model \textit{simultaneously}, using an end-to-end learning strategy.
We are able to achieve this thanks to the two properties of deep learning based reconstruction models: their speed and differentiable nature. These properties enable us to rapidly evaluate the effect of small changes to the sub-sampling pattern on reconstruction quality.

\subsection{Optimization of the Sub-sampling Pattern}
Some papers have proposed ways to optimize the sub-sampling pattern in compressed sensing MRI. The OEDIPUS framework~\cite{haldar2018oedipus} uses the information-theoretic Cramer-Rao bound to compute a deterministic sampling pattern that is tailored to the specific imaging context. Seeger et al~\cite{seeger2010optimization} present a  Bayesian approach to optimize k-space sampling trajectories under sparsity constraints.
The resulting algorithm is computationally expensive and does not scale well to large datasets. To alleviate this drawback, Liu et al.~\cite{liu2012under} propose a computationally more efficient strategy to optimize the under-sampling trajectory. However, this method does not consider a sophisticated reconstruction technique. Instead, they merely optimize for the simple method of inverse Fourier transform with zero-filling. 

Below, we describe the proposed method, LOUPE, that computes the optimal probabilistic sub-sampling mask together with a state-of-the-art rapid neural network based reconstruction model. We train LOUPE using an end-to-end unsupervised learning approach with retrospectively sub-sampled images.

\section{Method}

\subsection{Learning-based Optimization of the Under-sampling Pattern}

In this section, we describe the details of our novel problem formulation and the approach we implement to solve it. We call our algorithm LOUPE, which stands for Learning-based Optimization of the Under-sampling Pattern. LOUPE considers the two fundamental problems of compressed sensing simultaneously: the optimization of the under-sampling pattern and learning a reconstruction model that rapidly solves the ill-posed anti-aliasing problem.

In LOUPE, we seek a ``probabilistic mask''~$\vec{p}$ that describes an independent Bernoulli (binary) random variable $\mathcal{B}$ at each k-space (discrete Fourier domain) location on the full-resolution grid. Thus, a probabilistic mask $\vec{p}$ is an image of probability values in k-space. A binary mask $\vec{m}$ has a value of 1 (0) that indicates that a sample is (not) acquired at the corresponding k-space point. We assume $\vec{m}$ is a realization of $\vec{M} \sim \prod_{i} \mathcal{B}(\vec{p}_i)$, where $i$ is the k-space location index. Let $\vec{x}_j$ denote a full-resolution (e.g., 2D) MRI slice in the image (spatial) domain, where $j$ is the scan index. While $\vec{p}$, $\vec{M}$, $\vec{m}$ and $\vec{x}_j$ are defined on a 2D grid (in k-space or image domain), we vectorize them in our mathematical expressions. Our method is not constrained to 2D images and can be applied 3D sampling grids as well.

LOUPE aims to solve the following optimization problem:
\begin{eqnarray}
\arg \min_{\vec{p}, A} \mathbb{E}_{\vec{M} \sim \prod_{i} \mathcal{B}(\vec{p}_i)} \bigg[ \lambda \sum_{i}  \vec{M}_i + \sum_{j} \| A (F^{H} \textrm{diag}(\vec{M}) F \vec{x}_j) - \vec{x}_j \|_1\bigg], \label{eq:LOUPE1}
\end{eqnarray}
where $F$ is the (forward) Fourier transform matrix, $F^{H}$ is its inverse (i.e., Hermitian transpose of $F$), $A(\cdot)$ is an anti-aliasing (de-noising) function that we will parameterize via a neural network, $\vec{M}_i \sim \mathcal{B}(\vec{p}_i)$ is an independent Bernoulli, $\textrm{diag}(\vec{M})$ is a diagonal matrix with diagonal elements set to $\vec{M}$, $\lambda \in \mathbb{R}^{+}$ is a hyper-parameter, and $\| \cdot \|_1$ denotes the L1-norm of a vector. While in our experiments $\vec{x}_j$ is real-valued, $F$ and $F^{H}$ are complex valued, and $A(\cdot)$ accepts a complex-valued input. We design $A$ to output a real-valued image.

The first term in Eq.~\eqref{eq:LOUPE1} is a sparsity penalty that encourages the number of k-space points that will be sampled to be small. The hyper-parameter $\lambda$ controls the influence of the sparsity penalty, where higher values yield a more aggressive sub-sampling factor. We approximate the second term using a Monte Carlo approach. Thus the LOUPE optimization problem becomes:
\begin{eqnarray}
\arg \min_{\vec{p}, A} \lambda \sum_{i}  \vec{p}_i + \sum_{j} \frac{1}{K}\sum_{k=1}^K \| A (F^{H} \textrm{diag}(\vec{m}^{(k)}) F \vec{x}_j) - \vec{x}_j \|_1, \label{eq:LOUPE2}
\end{eqnarray}
where $\vec{m}^{(k)}$ is an independent binary mask realization of $\vec{M} \sim \prod_{i} \mathcal{B}(\vec{p}_i)$, and we use~$K$ samples.We further re-parameterize the second term of Eq.~\eqref{eq:LOUPE2}:
\begin{eqnarray}
\arg \min_{\vec{p}, A} \lambda \sum_{i}  \vec{p}_i + \sum_{j} \frac{1}{K}\sum_{k=1}^K \| A (F^{H} \textrm{diag}(\vec{u}^{(k)} \leq \vec{p}) F \vec{x}_j) - \vec{x}_j \|_1, \label{eq:LOUPE3}
\end{eqnarray}
where $\vec{u}^{(k)}$ is a realization of a random vector of independent uniform random variables on $[0,1]$, and $\vec{u}^{(k)} \leq \vec{p}$ is a binary random vector where each entry is set to 1 if the inequality is satisfied, and 0 otherwise.

\begin{figure}[th]
\includegraphics[width=\textwidth]{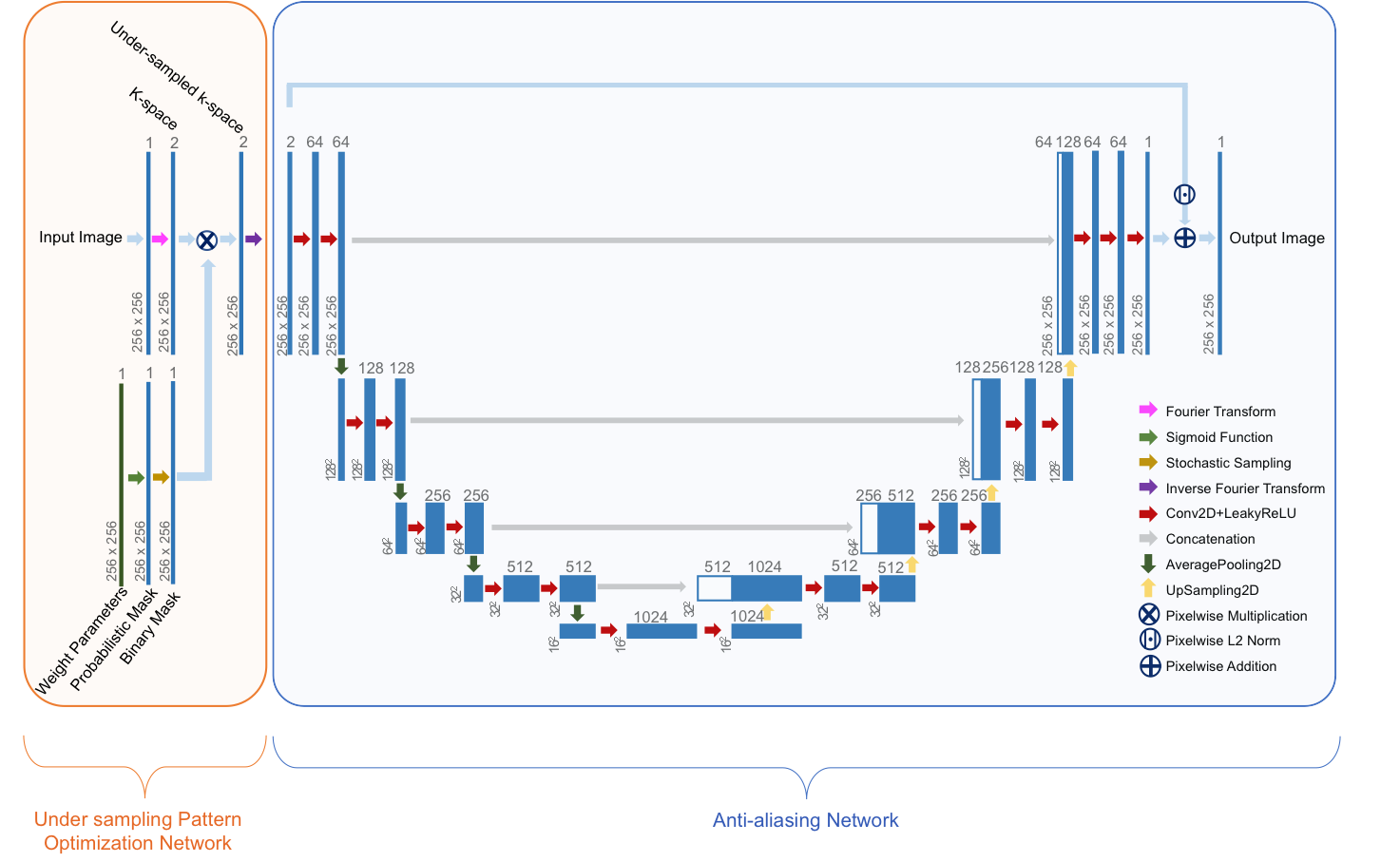}
\caption{The neural network architecture for LOUPE. Each vertical blue line represents a 2D image, with the number of channels indicated above and the size shown on the lower left side. The green line represents a 2D real-valued image of \textit{weight} parameters, where one parameter is learned at each location, which is then passed through a sigmoid to yield the probability mask~$\textbf{p}$.
} \label{fig:architecture}
\end{figure}
\subsection{Implementation}

We implement LOUPE using deep neural networks, which solve the learning problem via stochastic gradient descent.
To make the loss function differentiable everywhere, we relax the thresholding operation in Eq.~\eqref{eq:LOUPE3} via a sigmoid:
\begin{eqnarray}
\arg \min_{\vec{p}, \theta} \lambda \sum_{i}  \vec{p}_i + \sum_{j} \frac{1}{K}\sum_{k=1}^K \| A_{\theta} (F^{H} \textrm{diag}(\sigma_s(\vec{u}^{(k)}-\vec{p})) F \vec{x}_j) - \vec{x}_j \|_1, \label{eq:LOUPE4}
\end{eqnarray}
where $\sigma_s(a) = \frac{1}{1+e^{-sa}}$, and $A_\theta$ denotes a neural network parameterized with weights $\theta$. We set the slope for this sigmoid to be relatively steep to better approximate the thresholding step function.

The anti-aliasing function $A_{\theta}$ is a fully-convolutional neural network that builds on the widely used U-Net architecture~\cite{ronneberger2015u}. The input to $A_{\theta}$ is a two-channel 2D image, which correspond to the real and imaginary components. As in~\cite{lee2017deep}, the U-Net estimates the difference between the aliased reconstruction (i.e., the result of applying the inverse Fourier transform to the zero-filled under-sampled k-space measurements), and the fully-sampled ground truth image. Finally, the probabilistic mask $\vec{p}$ is formed by passing an unrestricted real-valued image through a sigmoid.
Figure~\ref{fig:architecture} illustrates the full architecture that combines all these elements. The red arrows represent 2D convolution layers with a kernel size $3 \times 3$, and a Leaky ReLU activation followed by Batch Normalization. The convolutions use zero-padding to match the input and output sizes. The gray arrows indicate skip connections, which correspond to concatenation operations. We also implement a stochastic sampling layer that draws uniform random vectors $\vec{u}^{(k)}$.
This is similar to the Monte Carlo strategy used in variational neural networks~\cite{kingma2013auto}.

We train our model on a collection of full-resolution images $\{\vec{x}_j\}$.
Thus, LOUPE minimizes the unsupervised loss function~\eqref{eq:LOUPE4} using an end-to-end learning strategy to obtain the probabilistic mask $\vec{p}$ and the weights $\theta$ of the anti-aliasing network $A_{\theta}$.
The hyper-parameter $\lambda$ is set empirically to obtain the desired sparsity.
We  implement our neural network in Keras~\cite{chollet2015keras}, with TensorFlow~\cite{abadi2016tensorflow} as the back-end and using layers from Neuron library ~\cite{dalca2018anatomical}. The code is made available at: ~\url{https://github.com/cagladbahadir/LOUPE}.
We use the ADAM~\cite{kingma2014adam} optimizer with an initial learning rate of 0.001 and terminate learning when validation loss plateaued. 
Our mini-batch size is 32 and $K=1$.
The input images are randomly shuffled.

\section{Empirical Analysis}
\subsection{Data}

In our analyses, we used 3D T1-weighted brain MRI scans from the multi-site ABIDE-1 study~\cite{di2014autism}.
We used 100 high quality volumes, as rated by independent experts via visual assessment, for training LOUPE, while a non-overlapping set of fifty subjects were used for validation.
For testing all methods, including LOUPE, we used ten  held-out independent test subjects.
All our experiments were conducted on 2D axial slices, which consisted of $1\times1 \textrm{mm}^2$ pixels and were of size $256 \times 256$.
We extracted 175 slices from each 3D volume, which provided full coverage of the brain - our central region of interest, and excluded slices that were mostly background. 

\subsection{Evaluation}

During testing, we computed peak signal to noise ratio (PSNR) between the reconstructions of the different models and the full-resolution ground truth images for each volume. PSNR is a standard metric of reconstruction quality used in compressed sensing MRI~\cite{sun2016deep}. Our quantitative results with other metrics (not included) were also consistent.

\subsection{Benchmark Reconstruction Methods}

The first benchmark method is ALOHA~\cite{lee2016acceleration}, which uses a low-rank Hankel matrix to impute missing k-space values. We employed the code distributed by the authors\footnote{\url{https://bispl.weebly.com/aloha-for-mr-recon.html}}. Since the default setting did not produce acceptable results on our data, we optimized the input parameters to minimize the MAE on a training subject.

The second benchmark reconstruction method we consider is a novel regularized regression technique that combines total generalized variation (TGV) and the shearlet transform. This method has been demonstrated to yield excellent accuracy in compressed sensing MRI~\cite{guo2014new}. We used the code provided by the authors\footnote{\url{http://www.math.ucla.edu/~wotaoyin/papers/tgv_shearlet.html}}.

Our third benchmark method is based on the Block Matching 3D (BM3D) method, which was recently shown to offer high quality reconstructions for under-sampled MRI data~\cite{eksioglu2016decoupled}. BM3D is an iterative method that alternates between a de-noising step and a reconstruction step. We employed the open source code\footnote{\url{http://web.itu.edu.tr/eksioglue/pubs/BM3D_MRI.htm}}.

Finally, we consider a U-Net based reconstruction method, similar to the recently proposed deep residual learning for anti-aliasing technique of~\cite{lee2017deep}. This reconstruction model is the one we used in LOUPE, with an important difference: in the benchmark implementation, the anti-aliasing model is trained from scratch, for each sub-sampling mask, separately. In LOUPE, this model is trained \textit{jointly} with the optimization of the sub-sampling mask.

\begin{figure}[t]
\includegraphics[width=\textwidth]{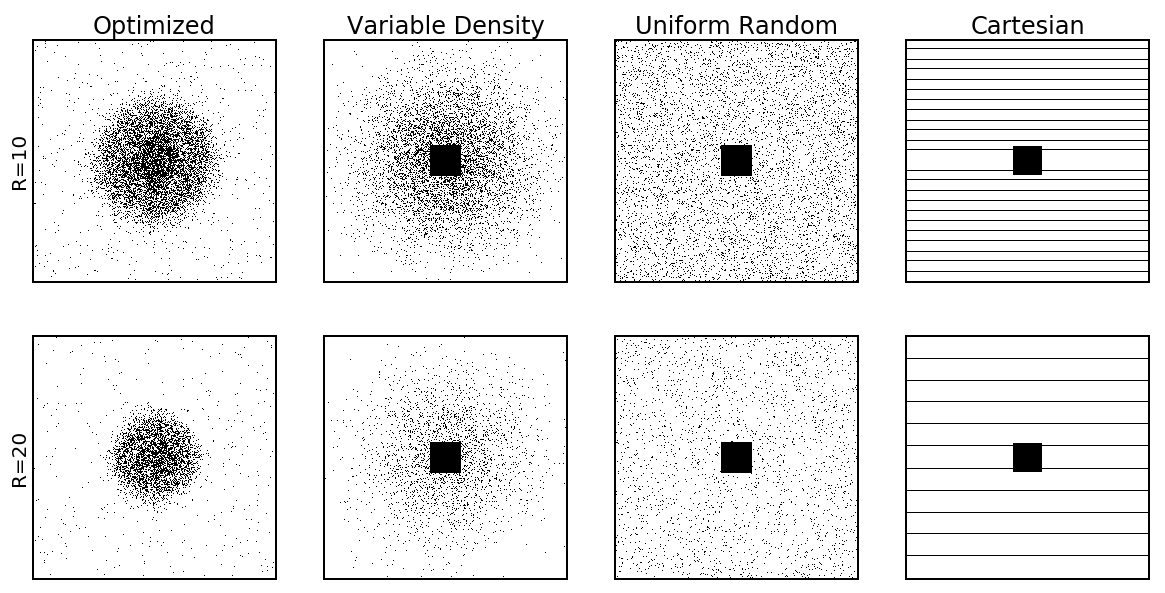}
\caption{Optimized and benchmark masks for two levels of sub-sampling rates: $R=10$ and $R=20$. Figures are in 2D k-space and black dots indicate the points at which a sample is acquired. Representative instantiations are visualized for the random masks.} \label{fig:masks}
\end{figure}

\subsection{Sub-sampling Masks}
In this study, we consider three different sub-sampling patterns that are widely used in the literature:
Random Uniform~\cite{gamper2008compressed}, Random Variable Density~\cite{wang2010variable} and equispaced Cartesian~\cite{haldar2011compressed} - all with a fixed $32 \times 32$ so-called ``calibration region'' in the center of the k-space. The calibration region is a fully sampled rectangular region around the origin, and has been demonstrated to yield better reconstruction performance~\cite{uecker2014espirit}. We experimented with excluding the calibration region and sub-sampling over the entire k-space. However, reconstruction performance was no better than including the calibration region, so we omit these results.

The Uniform and Variable Density patterns were randomly generated by drawing independent Bernoulli samples. For Uniform, the probability value at each k-space point was the same and equal to the desired sparsity level. For Variable Density, the probability value at each k-space point was chosen from a Gaussian distribution, centered at the k-space origin. The proportionality constant was set to achieve the desired sparsity level.
The Cartesian sub-sampling pattern is deterministic, and
yields a k-space trajectory that is straightforward to implement.
Figure~\ref{fig:masks} visualizes these masks.
We consider two sparsity levels: $10\%$ and $5\%$, which correspond to $R=10$ and $R=20$ sub-sampling rates.

\section{Results}

Table~\ref{tb:runtime} lists run time statistics for the different reconstruction methods, computed on the test subjects. For the U-Net, we provide run-times for both GPU (NVidia Titan Xp) and CPU. The U-Net model is significantly faster than the other reconstruction methods, which are all iterative. This speed, combined with the fact that the neural network model is differentiable, enabled us to use the U-Net in the end-to-end learning of LOUPE, and optimize the sub-sampling pattern.

Figure~\ref{fig:masks} shows the optimized sub-sampling mask that was computed by LOUPE on T1-weighted brain MRI scans from 100 training subjects. The resulting mask has similarities with to the Variable Density mask. While it does not include a calibration region, it exhibits a denser sampling pattern closer to the origin of k-space. However, at high frequency values, the relative density of the optimized mask is much smaller than the Variable Density mask.

\begin{figure}[t]
\includegraphics[width=\textwidth]{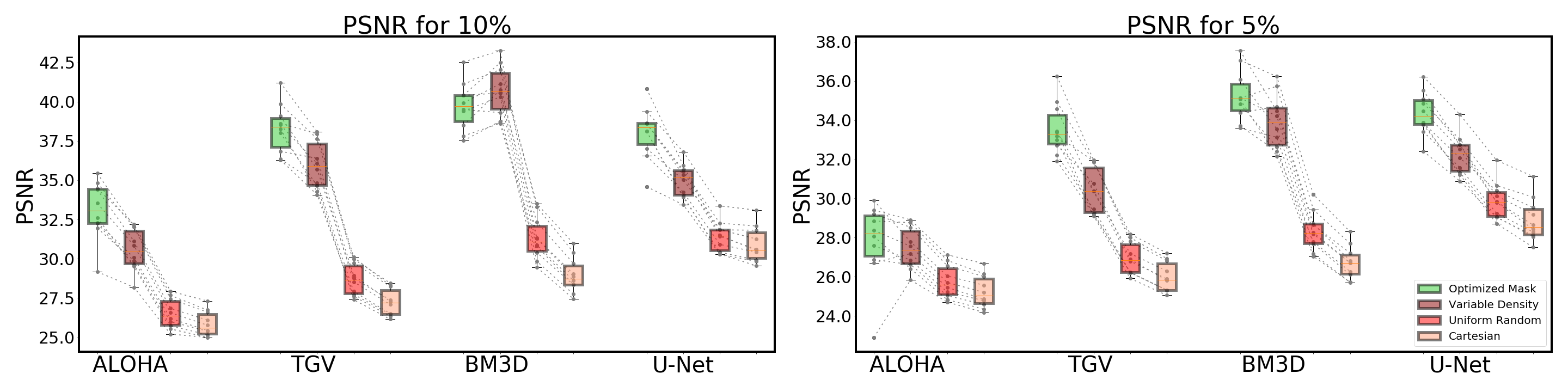}
\caption{Quantitative evaluation of reconstruction quality. For each plot, we show four reconstruction methods using four acquisition masks, including the Optimized Mask obtained using LOUPE in green. Each dot is the PSNR value for a single test subject across slices. For each box, the red line shows the median value, and the whiskers indicate the the most extreme (non-outlier) data points.} \label{fig:metrics}
\end{figure}

Figure~\ref{fig:metrics} includes box plots for subject-level PSNR values of reconstructions obtained with four reconstruction methods, four different masks, and two sub-sampling rates. The Cartesian and Uniform masks overall yielded worse reconstructions than the Variable Density and Optimized masks. In all except a single scenario, the Optimized mask significantly outperformed other masks (FDR corrected $q<0.01$ on paired t-tests). The only case where the Optimized mask was not the best performer was for the $10\%$ sub-sampling rate, coupled with the BM3D reconstruction method~\cite{eksioglu2016decoupled}. Here, the PSNR values were slightly worse than the best-performing mask, that of Variable Density.

\begin{table}[t]
	\centering
	\caption{Average per volume run times (in sec) for different reconstruction methods. All except U-Net (GPU) were evaluated on a CPU - a dual Intel Xeon (E5-2640, 2.4GHz).}
	\label{tb:runtime}
	\begin{tabular}{|c|c|c|c|c|}
		\hline
		ALOHA~\cite{lee2016acceleration} & TGV~\cite{guo2014new} & BM3D~\cite{eksioglu2016decoupled} &  U-Net~\cite{lee2017deep} (CPU) & U-Net~\cite{lee2017deep} (GPU) \\
		\hline
		$498 \pm 43.9$ & $492 \pm 33.8$ & $1691.1 \pm 216.4$ & $55.9 \pm 0.3$ & $1.6 \pm 0.4$ \\ \hline
	\end{tabular}
\end{table}

\begin{figure}
\includegraphics[width=\textwidth]{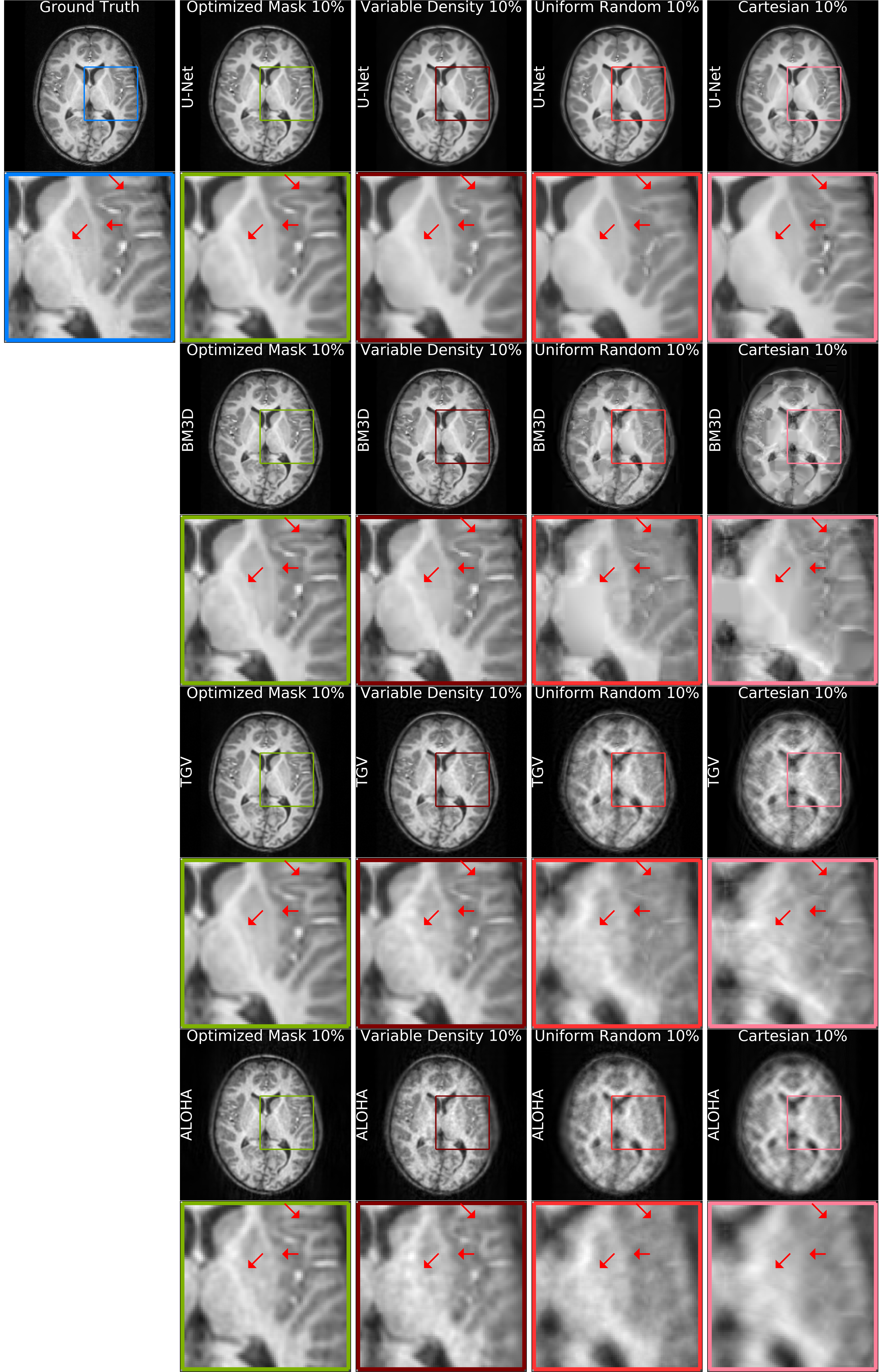}
\caption{Reconstructions for a representative slice and $10\%$ sub-sampling. Each row is a reconstruction method. Each column corresponds to a sub-sampling mask. We observe that our optimized mask yields reconstructions that capture more anatomical detail. Red arrows highlight some nuanced features that were often missed in reconstructions.} \label{fig:Brains1}
\end{figure}

\begin{figure}
\includegraphics[width=\textwidth]{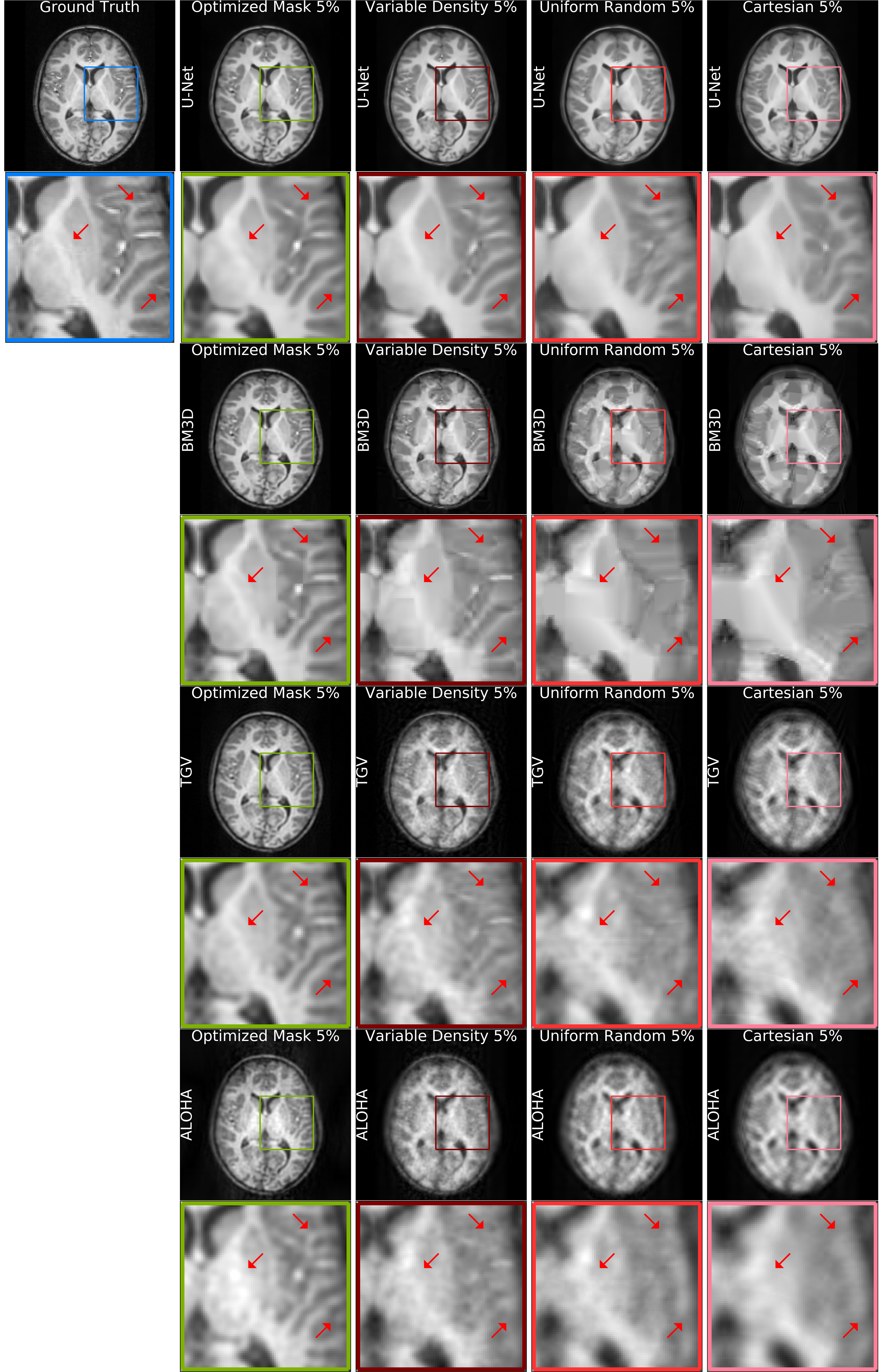}
\caption{Reconstructions for a representative slice and $5\%$ sub-sampling. See caption of Figure~\ref{fig:Brains1} and text for more detail.} \label{fig:Brains2}
\end{figure}

While the quantitative results give us a sense of overall quality, we found it very informative to visually inspect the reconstructions. Figures~\ref{fig:Brains1} and \ref{fig:Brains2} show typical examples of reconstructed images. We observe that our optimized mask yielded reconstructions that capture much more anatomical detail than what competing masks yielded (highlighted with red arrows in the pictures). In particular, the cortical folding pattern and the boundary of the putamen -- a subcortical structure -- were much better discernible for our optimized mask.
The difference in reconstruction quality between the different methods can also be appreciated. Overall, U-Net and BM3D offer more faithful reconstructions that can be recognized in the zoomed-in views.

\section{Discussion}

We presented a novel learning-based approach to simultaneously optimize the sub-sampling pattern and reconstruction model.
Our experiments on retrospectively under-sampled brain MRI scans suggest that our optimized mask can yield reconstructions that are of higher quality than those computed from other widely-used under-sampling masks.

There are several future directions we would like to explore.
First, sampling associated cost is captured with an L1 penalty in our formulation. We are interested in exploring alternate metrics that would better capture the true cost of a k-space trajectory, which is constrained by hardware limitations. Second, in LOUPE we used L1 norm for reconstruction loss. This can also be replaced with alternate metrics, such as those based on adversarial learning or emphasizing subtle yet important anatomical details and/or pathology. Third, we will consider combining LOUPE with a multi-coil parallel imaging approach to obtain even higher levels of acceleration. Fourth, we plan to explore optimizing sub-sampling patterns for other MRI sequences and organ domains.
More broadly, we believe that the proposed framework can be used in other compressed sensing and communication applications. 

\section*{Acknowledgements}
This work was supported by NIH R01 grants (R01LM012719 and R01AG053949), the NSF NeuroNex grant 1707312, and NSF CAREER grant (1748377).
\bibliographystyle{splncs04}
\bibliography{IPMI}

\end{document}